# WEP: AN ENERGY EFFICIENT PROTOCOL FOR CLUSTER BASED HETEROGENEOUS WIRELESS SENSOR NETWORK

Md. Golam Rashed[1], M. Hasnat Kabir[2], Shaikh Enayet Ullah[3]

[1]Department of Electronics and Telecommunication Engineering (ETE)
Prime University, Dhaka-1209, Bangladesh
golamrashed.ru@gmail.com

[2,3]Department of Information and Communication Engineering
University of Rajshahi, Rajshahi-6205, Bangladesh.
hasnatkabir@yahoo.com and enayet67@yahoo.com

*ABSTRACT*

*We develop an energy-efficient routing protocol in order to enhance the stability period of wireless sensor networks. This protocol is called weighted election protocol (WEP). It introduces a scheme to combine clustering strategy with chain routing algorithm for satisfy both energy and stable period constrains under heterogeneous environment in WSNs. Simulation results show that new one performs better than LEACH, SEP and HEARP in terms of stability period and network lifetime. It is also found that longer stability period strongly depend on higher values of extra energy during its heterogeneous settings.*

## 1. INTRODUCTION

A collection of mobile or static nodes which are able to communicate with each other for transferring data more efficiently and autonomously can be defined as wireless sensor network [1]. A lot of applications of wireless sensor network can be found in different field such as events, battlefield surveillance, recognition security, drug identification and automatic security etc. [2]. On going research on wireless sensor network is very active at present including numerous workshops and conferences arranged each year [3].

In wireless sensor network (WNS) one of the main constraints is limited battery power which plays a great influence on the lifetime and the quality of the network [4], [5]. Several routing protocols have been designed for wireless sensor networks to satisfy energy utilization and efficiency requirement [6] - [9]. Efficiency, scalability and lifetime of WSN can be enhanced using hierarchical routing. Here, sensors are organized themselves into clusters and each cluster has a cluster head. Currently, several energy efficient data collecting hierarchical protocols are available such as LEACH (Lower-Energy Adaptive Clustering Hierarchy), Stable Election Protocol (SEP) and Hierarchical Energy Aware Routing Protocol (HEARP) [9]-[11].

LEACH is a clustering protocol where cluster heads are randomly rotated to balance energy of network. The principle of LEACH is how to determine cluster and cluster head. The cluster head accepts data from other sensors within the same cluster, aggregate data and finally sends data to the Base Station (BS). Here all cluster heads are directly communicate to the BS. Since cluster heads are randomly choosing in LEACH algorithm so it has some probability to form a low-energy normal node as a cluster head. As a result, this cluster is not able to transfer data to the base station for long time. Therefore, network performance and lifetime will decrease. This problem was solved by Stable Election Protocol (SEP) [10]. SEP is a heterogeneous-aware protocol to prolong the time interval before the death of the first node. In this protocol, some sensors have high energy with respect to others therefore probability of these sensors will be increased to become as cluster heads. SEP successfully extends the stable region than LEACH. However, energy efficiency and low latency are considered as two key issues in designing a WSNs, routing protocol. To achieve these criterions, Hierarchical Energy Aware Routing Protocol (HEARP) [11] was proposed. It has been found from simulated results that HEARP is better than LEACH, in terms of energy consumption and latency. On the other hand, HEARP shows well balanced latency as compared to PEGASIS [12]. The main advantage of HEARP is increasing stability period of the WSNs and it saves energy because only one node transmits data directly to the base station. In this work we propose chain-based clustering routing protocols which overcome the instability problem of HEARP. Using chain among CHs instead of all nodes, we can alleviate the excessive delay to send data from sensing field to BS and achieve more energy gain.

## 2. THE NETWORK MODEL

We consider a 100-node network with randomly distributed nodes in a (100 x 100) meter area. The BS is located at (x=50, y=50). The snapshot of the considered network is shown in Figure 1. The length of each signal is 4000 bits and the energy required for data aggregation is 5nJ/bit/signal. Moreover, the initial energy of each node is 0.1 *joule*.

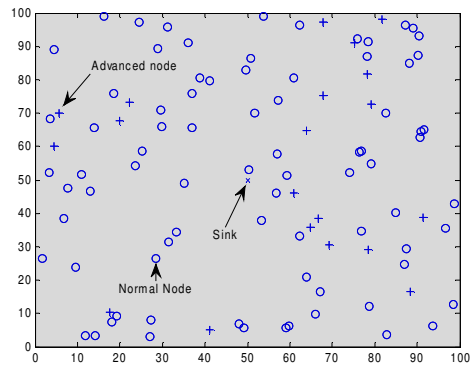

Figure 1. A wireless sensor network

In the paper, we have considered the following assumptions:
- Each sensor node has power control and the ability to transmit data to any other sensor node or directly to the BS.
- Our model is based on the clustering hierarchy process using the characteristic parameters of heterogeneity, namely the fraction of advanced nodes (m) and the additional energy factor (α) between advanced and normal nodes.
- Advanced nodes have to become cluster heads more often than that of normal nodes by separate threshold for each type of nodes.
- There is no mobility.

In our work, we use the first order radio model as describe in LEACH.

## 3. OUR PROPOSED APPROACH

It has been seen from existing hierarchical routing protocols that most of them ensure longer life time of the network. However, one of the major challenges in wireless sensor network is to

55

prolong the time interval before the death of first node. It can be referred as *Stability period*. Without longer stability period, more information could not be able to collect from the sensor field even though the life time of the network is high. So prolonging the stability period is crucial for many applications.

To meet the need to prolong the stability period of the network we propose a scheme to combine clustering strategy with chain routing algorithm in order to satisfy both energy and stable period constraints under heterogeneous environment in WSNs. The nodes in sensor field do not hold same energy called heterogeneous setting. Here two types of nodes are used such as advanced node and normal node where advanced nodes have more energy than normal ones. Clustering strategy is useful for low latency while chain routing algorithm is beneficial for energy efficiency thereby longer stability period can be achieved. In order to prolong the stable region, our proposed approach attempts to maintain the constraint of well balanced energy consumption. Intuitively, advanced nodes have to become cluster heads more often than that of normal nodes.

Suppose that $E_0$ is the initial energy of each normal sensor. The energy of each advanced node is then $E_0(1+\alpha)$ if we consider that advanced nodes should have α time more energy than normal nodes and M(%) nodes will be equipped with additional energy. Considering these assumptions in network model, the total (initial) energy of the new heterogeneous setting is equal to:

$$n \cdot (1-m) \cdot E_0 + n \cdot m \cdot E_0 \cdot (1+\alpha) = n \cdot E_0 \cdot (1+\alpha*m)$$

So, the total energy of the system is increased by a factor of $1+\alpha*m$. Therefore, the number of epoch is proportional to the increment of energy. The new epoch must become equal to $\frac{1}{p_{opt}}*(1+\alpha*m)$ because the system has α*m times more energy and virtually α*m more nodes (with the same energy as the normal nodes). The desire number of new epoch using the same number of nodes (advanced, normal) can be achieved by increasing the probability of the advanced nodes to be elected as cluster head more than once per epoch. Here "epoch" means the total number of rounds by which all the sensor nodes become a cluster head once.

Our approach is to assign a weight to the optimal probability $p_{opt}$. This weight must be equal to the initial energy of each node divided by the initial energy of the normal node. After assigning weighted probability of each type nodes, we can select cluster head and their associated non-cluster head as the same way as it done in LEACH protocol. Then we can use greedy algorithm to make a chain among the selected cluster heads. After constructing chain among cluster head nodes a chain leader is selected randomly. Using TDMA schedule, all non-cluster head nodes send their data to their respective cluster head nodes. The cluster head nodes in each cluster then fused those data and finally send to the base station. We use MATLAB [13] to simulate our new approach together with three other routing protocols, namely LEACH, SEP, HEARP. The simulation result of our approach will compare with them.

## 4. SIMULATION RESULT AND ANALYSIS

To evaluate the performance of our proposed approach, we vary the heterogeneous parameters (α, m) and investigate the improvement of stability period and network lifetime with each of the variations. Furthermore we analyze the total energy utilization of the network.

Figure 2 shows the result for the case of $m$=0.2 and $\alpha$=3. It can be seen from this figure that the stable region of proposed system moderately changed with respect to other protocol. The stable region of our proposed approach is increased significantly by 13% from LEACH, 6% from SEP, and 8% from HEARP.

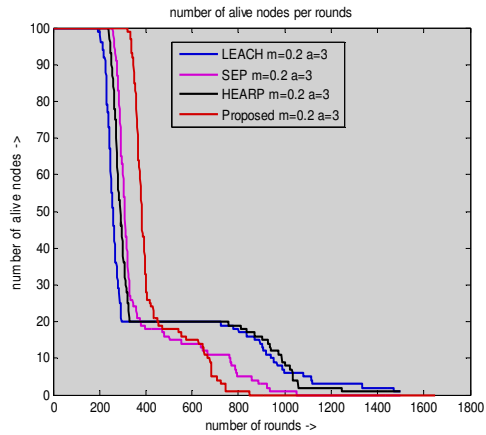

Figure 2. Comparison among LEACH, SEP, HEARP, and proposed approach (WEP) in the presence of heterogeneity (m=0.2 and $\alpha$=3).

Figure3 and 4 show the summarized improvement both in stable and unstable region of our approach over LEACH, SEP, and HEARP, respectively. It can be pointed out that the stable and unstable region strongly depends on heterogeneous parameters.

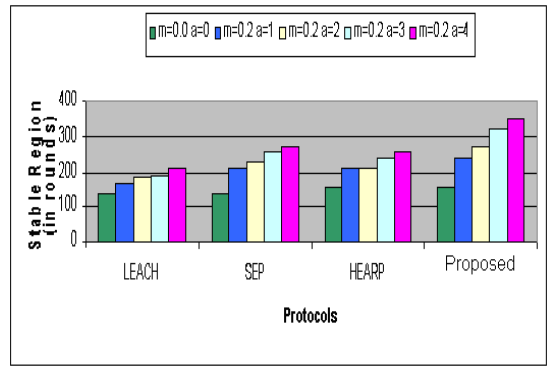

Figure 3. Length of Stable region in rounds for different values of heterogeneity

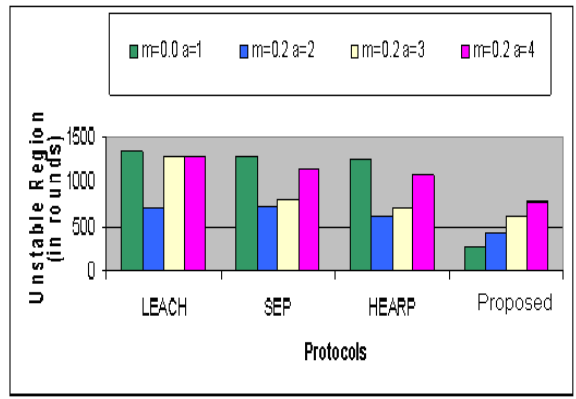

Figure 4. Length of Unstable regions in rounds for different values of heterogeneity

On the other hand, figure 4 shows the comparative results of unstable region of different protocols in terms of rounds. It is observed that the unstable region of WEP is significantly less than other protocols. The reason can be explain as the weighted probability of electing cluster heads is proportional to initial energy of the nodes. We have found that presented protocol increases stability period for higher values of extra energy come from more powerful nodes (Advanced nodes).

It is important to know when the first node will die which indicates the stability period. The metric *first of the Nodes Dies* (FND) denotes an estimated value for a specific network configuration. Furthermore, sensors can be placed in proximity to each other. Thus, adjacent sensors could record related or identical data. Hence, the loss of a single or multiple nodes does not automatically diminish the quality of service of the network. In this case the metric *Half of the Nodes Dies* (HND) indicates an estimated value for the half-life period of a micro-sensor network. Finally, the metric *Last Node Dies* (LND) provides an projected value for the overall lifetime of a micro-sensor network.

For a cluster-based algorithm like LEACH, SEP, HEARP, the metric LND is not interesting since more than one node is necessary to perform the clustering algorithm. Hence, we discuss our result related to network life time in this paper in the light of FND and HND metrics. Fig. 5 illustrates lifetime of the sample network with our proposed protocols.

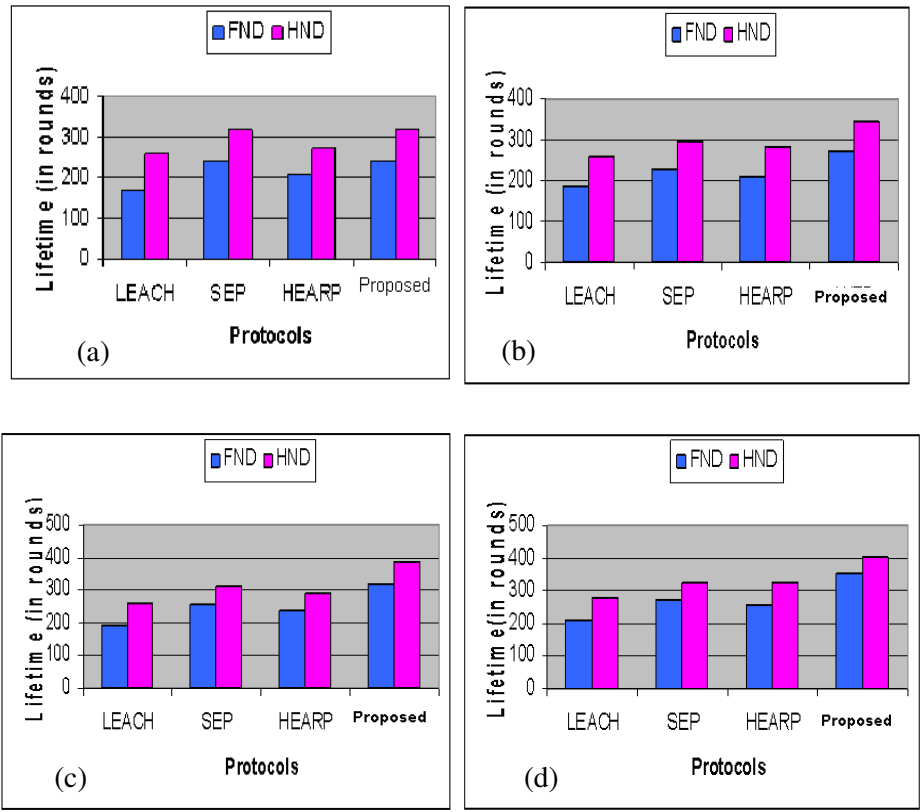

Figure 5. Comparison of lifetime for WEP and other clustered protocols with different heterogeneous case (a) *m*=0.2 and $\alpha$=1, (b) *m*=0.2 and $\alpha$=2, (c) *m*=0.2 and $\alpha$=3, (d) *m*=0.2 and $\alpha$=4.

Since the nodes have limited energy and it is used during the course of network operation. Certain energy is reduced whenever a node transmits or receives data and whenever it performs data fusion. Once a node runs out of its energy, it is considered to be dead and it can no longer transmit or receive any data. The simulation ends when all the nodes in the network run out of their energy. High energy efficiency means low energy consumption and long stability period of the micro-sensor network. From the above figures we can found that the lifetime (in round) increases in WEP. It is interesting to point out that the first node dies after twice times later than that of LEACH and it is strongly depend on $\alpha$. Figure 6 demonstrates the comparative total energy consumption of LEACH, SEP, HEARP, and our approach during stable period and unstable period for the heterogeneous settings m=0.2 and $\alpha$=3. According to network model stated in section 2.1, if initial energy of each node is 0.1 joules then total amount of energy of the network under the considered heterogeneous setting is about 16 joules. It is expected that maximum energy should be consumed during stable period than unstable period because in stable period all nodes will be engaged to transmit or receive data and to perform data fusion thereby consuming more energy.

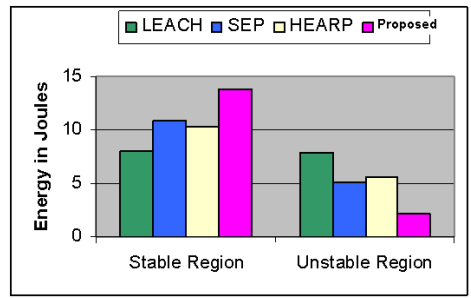

Figure 6: Comparison of total energy consumption in LEACH, SEP, HEARP, and WEP during stable region and unstable region for the heterogeneous settings of m=0.2 and $\alpha$=3.

From figure 6 it is obvious that our proposed approach utilizes maximum amount by 86.30% of total network energy during stable region where as LEACH (by 50%), SEP (by 63.25%), HEARP (by 66 %).

## 5. CONCLUSION

Our proposed approach gently improvements the stable period from LEACH, SEP, and HEARP by reducing the unstable period of the networks which is essential for some specific applications. The simulated results show that the first node dies after certain rounds later than that of others protocol and it strongly depends on the value of $\alpha$. However, WEP utilizes 86.3% of total energy during stable region. Therefore, it can be concluded that our proposed protocol provides energy efficient routing that ensures longer stability period.

## ACKNOWLEDGEMENTS

This paper is supported by Ministry of Science, Information & Communication Engineering (MOSICT), of the Government of the People's Republic of Bangladesh.

**Authors**

**Md. Golam Rashed** received the B.Sc. and M.Sc. degree in Information and Communication Engineering from University of Rajshahi, Bangladesh in 2008 and 2009, respectively. He is currently working as a lecturer in Prime University, Bangladesh.

**M. Hasnat Kabir** received the B.Sc. and M.Sc. degree in Applied Physics and Electronics from University of Rajshahi, Bangladesh in 1995 and 1996, respectively. He has completed his PhD from Kochi University of Technology, Japan in 2007. Now he is working as an Assistant Professor in the dept. of Information and Communication Engineering, University of Rajshahi, Bangladesh.

**Shaikh Enayet Ullah** received the B.Sc. and M.Sc. degree in Applied Physics and Electronics from University of Rajshahi, Bangladesh. He has completed his PhD from Jahangirnagor University, Bangladesh. He is currently working as a Professor and Chairman of the dept. of Information and Communication Engineering, University of Rajshahi, Bangladesh.